\documentclass[11pt,twocolumn,superscriptaddress]{revtex4}   
\pdfoutput=1
\usepackage{amsmath}
\usepackage{graphicx}
\usepackage{natbib}
\usepackage{mathrsfs}

\usepackage[T1]{fontenc}
\usepackage[letterpaper,textwidth=7in,top=.75in,bottom=.75in]{geometry}
\begin{document}
\title{Optical properties of the nitrogen-vacancy singlet levels in diamond}

 \author{V.~M.~Acosta}
 \email{vmacosta@berkeley.edu}
    \address{
     Department of Physics, University of California,
     Berkeley, CA 94720-7300
    }

 \author{A.~Jarmola}
    \address{
     Department of Physics, University of California,
     Berkeley, CA 94720-7300
    }

 \author{E.~Bauch}
    \address{
     Department of Physics, University of California,
     Berkeley, CA 94720-7300
    }

 \author{D.~Budker}
    \address{
     Department of Physics, University of California,
     Berkeley, CA 94720-7300
    }
    \address{Nuclear Science Division, Lawrence Berkeley National Laboratory,
     Berkeley CA 94720, USA
    }
\date{\today}

\begin{abstract}
We report measurements of the optical properties of the 1042 nm transition of negatively-charged Nitrogen-Vacancy (NV) centers in type 1b diamond. The results indicate that the upper level of this transition couples to the $m_s=\pm1$ sublevels of the ${^3}E$ excited state and is short-lived, with a lifetime of $\lesssim1~{\rm ns}$. The lower level is shown to have a temperature-dependent lifetime of $462(10)~{\rm ns}$ at 4.4 K and $219(3)~{\rm ns}$ at 295 K. The light-polarization dependence of 1042 nm absorption confirms that the transition is between orbitals of $A_1$ and $E$ character. The results shed new light on the NV level structure and optical pumping mechanism.
\end{abstract}
\maketitle

The negatively-charged Nitrogen-Vacancy (NV) center in diamond has received considerable attention in recent years for its unique optical and spin properties. The center has a paramagnetic ground state with long coherence times ($\gtrsim1~{\rm ms}$ \cite{BAL2009}), and spin levels can be initialized and detected by optical excitation in a broad range of wavelengths ($450\mbox{-}650~{\rm nm}$) \cite{DAV1974,RED1991,HE1993A,GRU1997} and manipulated on sub-nanosecond timescales \cite{FUC2009} by microwave (MW) excitation, all at room temperature. These traits, coupled with the excellent thermal, electrical, and mechanical properties of the diamond host, make NV centers promising candidates for quantum computing \cite{JEL2004,CHI2006,DUT2007,NEU2008,NEU2010,TOG2010} and ultra-sensitive metrology \cite{TAY2008,BAL2008,MAZ2008NATURE,ACO2009,MAE2010,STE2010}.

The NV center is now among the most widely studied defects in a solid \cite{JEL2006}, but there still exist open questions regarding its electronic level structure, and the mechanism by which optical excitation produces high spin polarization ($\gtrsim80\%$ even at room temperature) \cite{HAR2006,FEL2009,NEU2010} is not fully understood. The ordering and relative energies of all four singlet levels predicted by group theory \cite{LEN1996} have not been experimentally verified, and current theories disagree \cite{MAN2006,DEL2010,MA2010}. Since some of these levels are responsible for the optical pumping mechanism of the NV center, a full characterization of their properties is necessary before applications can reach their full potential.

Recently an infrared (IR) zero-phonon line (ZPL) at $\sim1042~{\rm nm}$ was observed after optical excitation at 532 nm \cite{ROG2008,IRNOTE}. The transition was characterized as a singlet-singlet transition involved in the optical pumping mechanism, but attempts to excite it on the phonon sideband were unsuccessful. In this Letter, we study the 1042 nm transition using laser light tuned to the ZPL. We determine the lifetimes of both singlet levels involved in the transition and infer their symmetry character from observed light-polarization selection rules. The results confirm the role of the transition in optical pumping of the center and give new information on the relative ordering and energies of the singlet levels.

\begin{figure}
\centering
    \includegraphics[width=.35\textwidth]{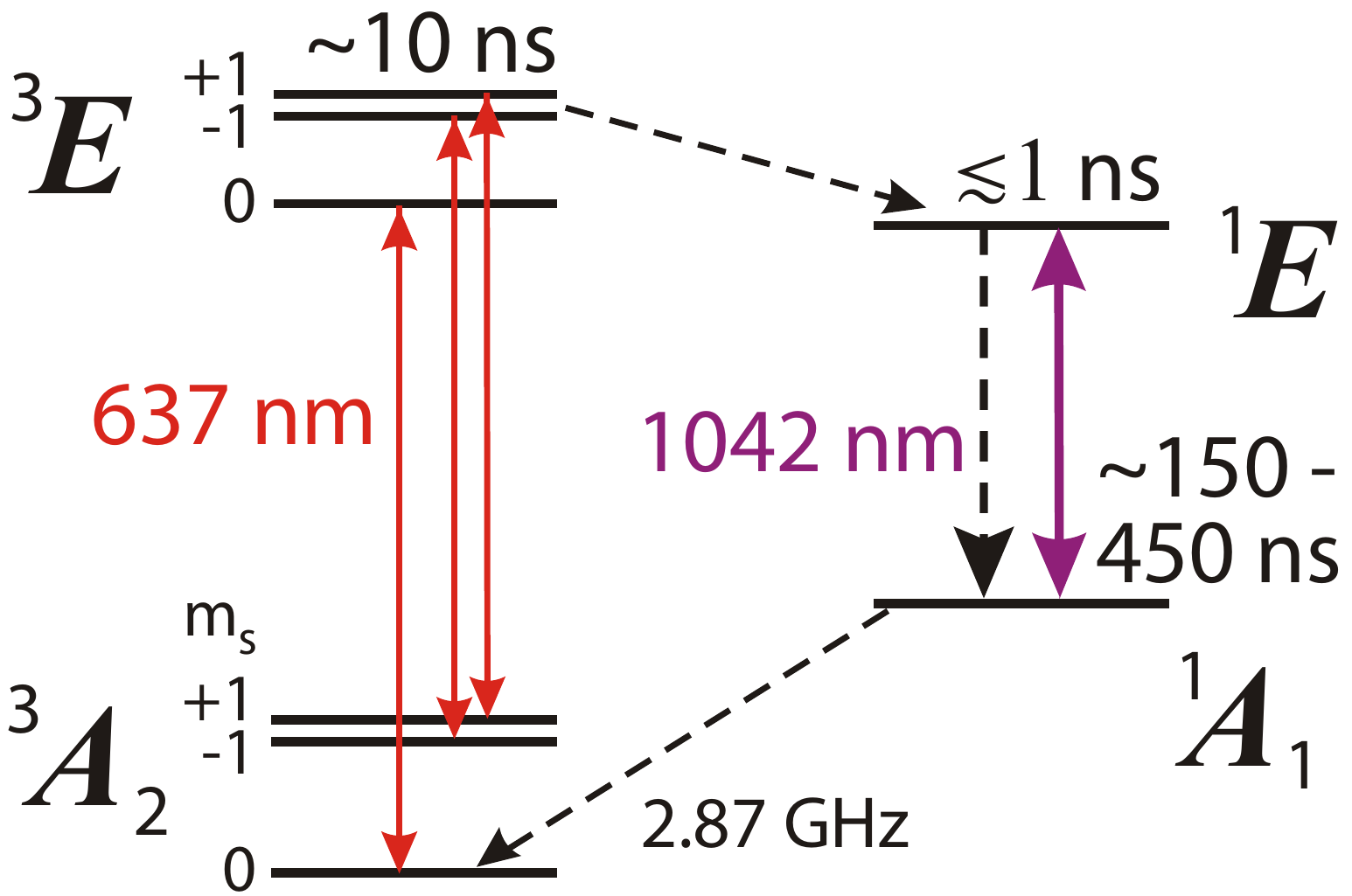}
    \caption{\footnotesize \label{fig:structure}Level diagram for NV center showing spin-triplet ground and excited states, as well as the singlet system involved in intersystem crossing. Radiative transitions are indicated by solid arrows and non-radiative transitions by dashed arrows. The tentative label of the upper(lower) singlet as ${^1}E$(${^1}A_1$) is based on the observed spin-selective decay paths (see text).}
\end{figure}

Figure \ref{fig:structure} depicts the energy level diagram of the NV center. Only experimentally verified levels are shown, though two additional singlet levels have been theoretically predicted \cite{LEN1996,MAN2006}. Transitions from the ${^3}A_2$ ground state to the ${^3}E$ state can be excited optically with ZPL at 637 nm. At least one singlet level lies close in energy to ${^3}E$, and spin-orbit coupling induces triplet-singlet intersystem crossing (ISC). NV centers in the $m_s=\pm1$ magnetic sublevels have significantly higher probability to undergo ISC \cite{HE1993A,GRU1997}, and, from symmetry considerations, this suggests that the closest lying singlet is ${^1}E$ \cite{MAN2006,MAN2007,MA2010}. NV centers which undergo ISC then decay to another, longer-lived singlet level \cite{ROG2008} (ZPL at 1042 nm), after which they cross over predominately to the $m_s=0$ sublevel of the ${^3}A_2$ ground state \cite{MAN2006,MA2010}. Symmetry considerations predict that this metastable singlet (MS) is ${^1}A_1$ \cite{MAN2006}, but recent theoretical calculations \cite{DEL2010,DOH2010} contradict this assignment.

\begin{figure}
\centering
    \includegraphics[width=.45\textwidth]{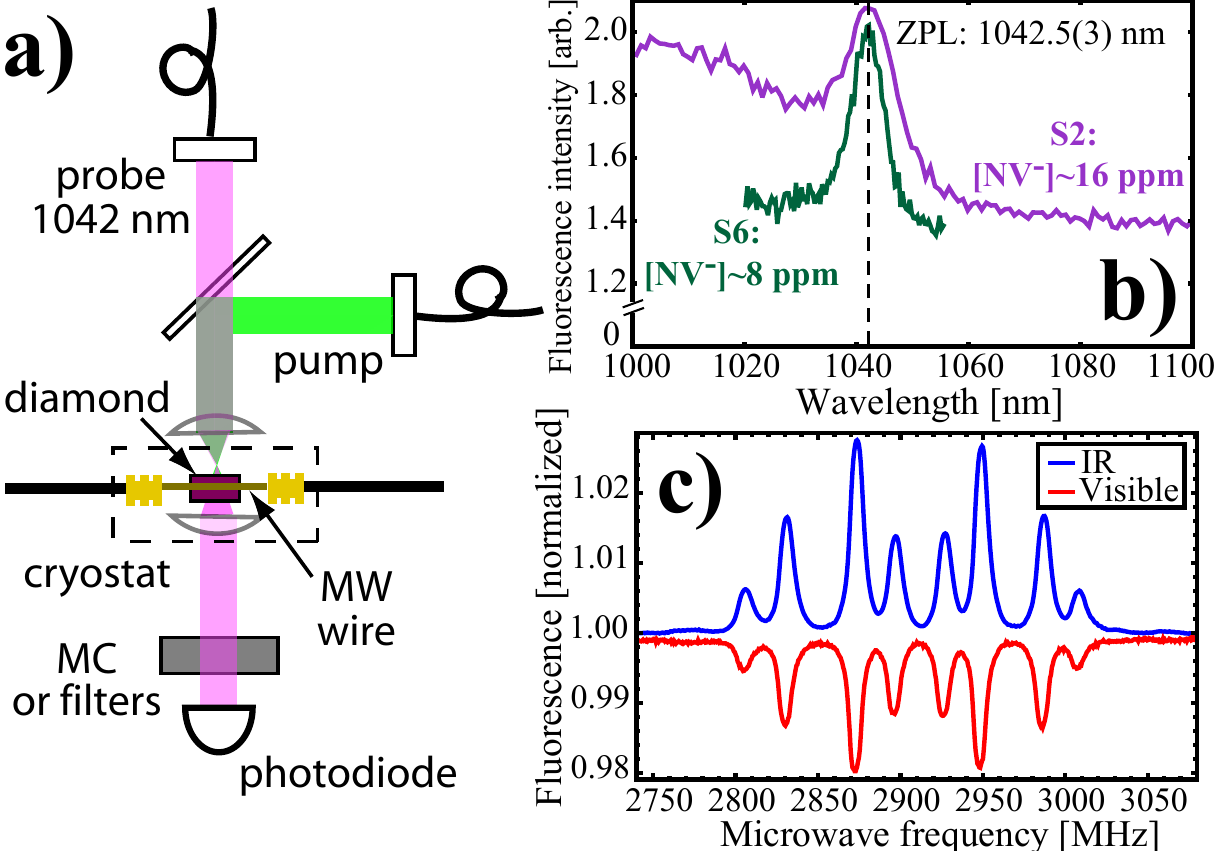}
    \caption{\footnotesize \label{fig:app} (a) Apparatus used for magnetic resonance and optical studies. MC--monochromator. (b) Fluorescence spectrum of the IR transition (excitation at 532 nm) at room temperature for two diamonds with different NV$^{\mbox{-}}$ concentrations \cite{ACO2009} showing a ZPL at 1042.5(3) nm. The sloped background is due to the tail of the phonon sideband of the much-stronger ${^3}E\rightarrow{^3}A_2$ transition \cite{ROG2008}. (c) Magnetic-resonance spectra of sample S2 for visible ($650\mbox{-}800~{\rm nm}$ bandpass filter) and IR (1000 nm longpass filter) emission at room temperature under an applied field of $\sim4~{\rm mT}$. Excitation was with $\sim1~{\rm W}$ of 532 nm light, polarized in the plane of the sample surface. The variation in peak heights can be explained by the pump-light and microwave polarization selection rules \cite{ALE2007}.}
\end{figure}

Figure \ref{fig:app}(a) illustrates the experimental apparatus used for optical characterization of the 1042 nm transition. Two laser beams, a pump (514, 532, or 637 nm) and IR probe ($\sim10~{\rm mW}$ at 1042.4 nm), were overlapped and focused with a 1/2"-focal-length lens onto the surface of a diamond housed in a liquid-helium cryostat (Janis ST-500). The temperature was measured by a diode located at the base of the cryostat's sample holder. The diamonds used in this work \cite{ACO2009}, grown under high-pressure-high-temperature conditions (HPHT), were irradiated with 3-MeV electrons and annealed at 750 C for two hours, after which they contain $\sim8\mbox{-}16~{\rm ppm}$ of NV$^{\mbox{-}}$ and $\sim50~{\rm ppm}$ of unconverted nitrogen. For absorption measurements, the transmission of the probe beam was spectrally filtered and detected with a Si photodiode. The same detection scheme was employed for fluorescence measurements, except the probe beam was blocked and different spectral filters were used, as appropriate. Figure \ref{fig:app}(b) shows the room-temperature fluorescence spectra for two different high-NV-density HPHT samples, S2 and S6, recorded using a monochromator. For the samples and temperature range ($\sim4.4\mbox{-}300~{\rm K}$) explored here, we measure the ZPL to be at 1042.5(3) nm. The IR emission is predominately concentrated in the ZPL \cite{ROG2008}, but the integrated intensity is less than visible emission (650-800 nm) by a factor of around 1000, presumably due to competition with a non-radiative decay process \cite{ROG2008,MA2010}.

Here we describe measurements made on S2. Figure \ref{fig:app}(c) shows room-temperature magnetic-resonance spectra detected in visible and IR emission with a 532 nm pump beam. These spectra can be understood as follows. Optical pumping collects NV centers in the $m_s=0$ ground-state sublevel. When the frequency of an applied MW field coincides with one of the eight ground-state magnetic-resonance transition frequencies (see, for example, Refs. \cite{ALE2007,ACO2009,LAI2009,MAE2010,STE2010,ACO2010}), population is transferred to an $m_s=\pm1$ sublevel. Under continuous optical excitation, this results in more NV centers undergoing ISC to the singlet levels, resulting in a decrease in visible fluorescence and an increase in IR fluorescence. Due to a combination of background fluorescence from the phonon sideband of the ${^3}E\rightarrow{^3}A_2$ transition, imperfect spin polarization, and non-zero branching ratio from $m_s=0$, the contrast of the IR fluorescence magnetic-resonance spectra is only a few percent.

To verify the model in Fig. \ref{fig:structure}, wherein the upper singlet is populated indirectly by decay from the $m_s=\pm1$ sublevels of ${^3}E$, we observed, via fluorescence, magnetic resonance spectra while pumping resonantly (637 nm) and non-resonantly (514 nm) at 70 K. The pump intensities were kept relatively low ($\lesssim1~{\rm kW/cm^2}$) in order to minimize bleaching \cite{DRA1999,HAN2010}. Under both pumping conditions, we found that the intensity ratio of visible emission (650-800 nm) to infrared emission (1000+ nm) was the same ($3(1)\times10^{3}$), confirming that the upper singlet is populated via the ${^3}E$. Further, the shape and contrast of the IR fluorescence spectra were similar under both pumping conditions, indicating that decay is primarily from the $m_s=\pm1$ sublevels.

\begin{figure}
\centering
    \includegraphics[width=.45\textwidth]{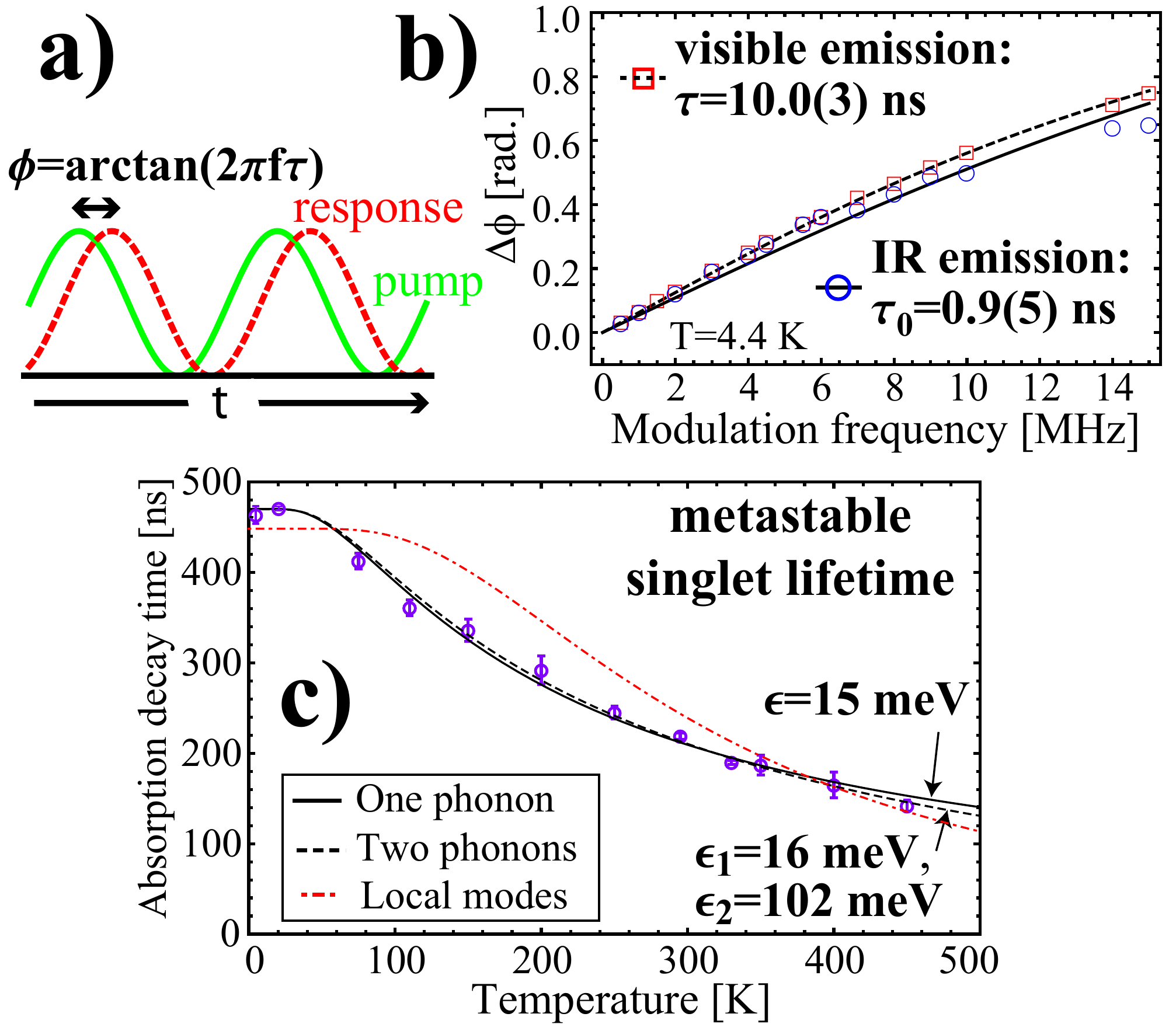}
    \caption{\footnotesize \label{fig:lifetimes} (a) The phase shift technique. (b) Phase shift of visible (650-800 nm) and IR (1035-1045 nm) fluorescence as a function of modulation frequency and fits (see text). (c) Lifetime of the MS as a function of temperature as well as fits to Eq. \eqref{eq:phonon}. The values of $\epsilon$ and $\epsilon_{1,2}$ are the fitted phonon energies for $N=1$ and $N=2$ decays, respectively. Error bars represent the standard deviation of multiple measurements. }
\end{figure}

In order to determine lifetimes of the upper singlet and ${^3}E$ excited state, fluorescence lifetime measurements were carried out for both the IR and visible emission using the phase-shift technique \cite{LAK2006} (Fig. \ref{fig:lifetimes}(a)). When the intensity of the pump beam is modulated sinusoidally with frequency $f$, the time-dependent fluorescence oscillates with a relative phase shift, $\phi$. For decay described by a single time-constant, $\tau$, which was used to model the visible emission \cite{VISNOTE2010}, this phase shift is $\phi=\arctan(2\pi f \tau)$. For the cascade decay expected for IR fluorescence (Fig. \ref{fig:structure}), we assume that the upper singlet level (lifetime, $\tau_0$) is populated only by the $m_s=\pm1$ sublevels in $^{3}E$ (lifetime, $\tau_1\approx7.8~{\rm ns}$ \cite{BAT2008}), and find $\phi=\arctan(\frac{2\pi f(\tau_1+\tau_0)}{1-(2\pi f)^2\tau_1\tau_0})$.

Figure \ref{fig:lifetimes}(b) shows results of the lifetime measurements at zero magnetic field for visible and IR emission, respectively. The pump (514 nm) modulation-frequency was varied and both the pump and fluorescence intensities were simultaneously recorded on separate detectors. The relative phase was determined by fitting both signals and correcting for time delays due to the detection scheme. For visible emission, the ensemble-averaged decay time is found to be $10.0(3)~{\rm ns}$. For IR emission, the fitted lifetime of the upper singlet is $\tau_0=0.9(5)~{\rm ns}$, which is consistent with the decay being dominated by a non-radiative channel \cite{ROG2008}. No temperature dependence was observed for either visible or IR fluorescence lifetimes in the range explored here ($4.4\mbox{-}70~{\rm K})$.

The lifetime of the MS level was measured by recording the transmission of 1042.4 nm light after abruptly turning the pump beam off (fall time $\lesssim20~{\rm ns}$), and fitting to a single exponential. Laser-intensity-dependent effects were determined to be smaller than the statistical uncertainty by varying both pump and probe laser powers. Figure \ref{fig:lifetimes}(c) shows the lifetime as a function of temperature, along with fits to the model described below. A substantial temperature dependence is observed; the MS lifetime is determined to be $462(10)~{\rm ns}$ at 4.4 K and decreases to $219(3)~{\rm ns}$ at 295 K and $142(6)~{\rm ns}$ at 450 K. The timescale of this decay is in agreement with the observed timescale for ISC to the ground state \cite{JEL2006,MAN2006}, confirming that the lower-energy level of the 1042 nm transition is indeed the metastable state which governs spin dynamics under optical excitation.

As the spontaneous emission rate is independent of temperature, a likely mechanism for temperature-dependent decay is emission stimulated by lattice phonons. Assuming a simple model where decay is dominated by a single channel, the temperature-dependent rate for the $N$-phonon decay is then given by \cite{PAR1967}:
\begin{equation}
\label{eq:phonon}
\frac{1}{\tau[T]}=\frac{1}{\tau[0]}\prod_{i=1}^N(1+n_i[T]),
\end{equation}
where $n_i[T]=(e^{\epsilon_i/(k_{\rm B} T)}-1)^{-1}$ is the Bose occupancy for the phonon mode with energy $\epsilon_i$, and $k_{\rm B}$ is the Boltzmann constant. Figure \ref{fig:lifetimes}(c) shows fits of Eq. \eqref{eq:phonon} for $N=1$ and 2. The fit parameters indicate that the total energy gap between the MS and final state, $\sum_i\epsilon_i$, is $15(1)~{\rm meV}$ for $N=1$ and $118(30)~{\rm meV}$ for $N=2$. We also fit the data to the dominant local phonon modes \cite{ROG2008}, modifying Eq. \eqref{eq:phonon} as $1/\tau[T]=1/\tau[0]\prod_{i=1}^2(1+n_i[T])^{N[\epsilon_i]}$, where $\epsilon_i=\{43,137\}~{\rm meV}$ are the phonon energies, and $N[\epsilon_i]$ are the respective degeneracies. While the fit is not as good, it suggests a similarly small energy gap of $129~{\rm meV}$ ($N[43~{\rm meV}]=3$ and $N[137~{\rm meV}]=0$). Such a small energy gap is not predicted by current models, where the MS is expected to lie $\sim700~{\rm meV}$ above the ${^3}A_2$ ground state \cite{ROG2008,DEL2010,MA2010}, so further work is necessary to resolve this.



\begin{figure}
\centering
    \includegraphics[width=.45\textwidth]{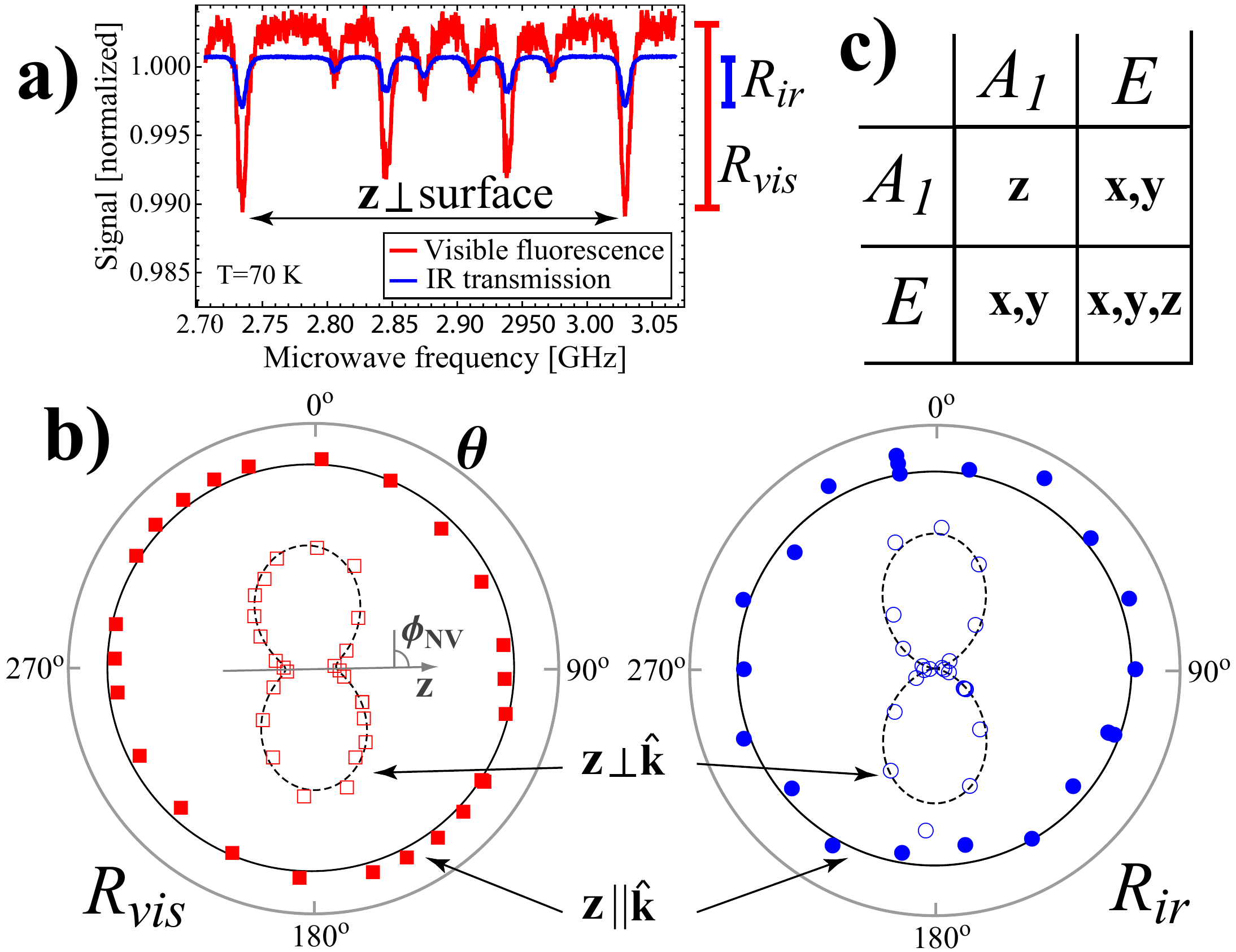}
    \caption{\footnotesize \label{fig:pol} (a) Example magnetic-resonance spectrum for visible fluorescence and transmission of IR probe at 70 K. As the sample was [111]-oriented, one of the NV axes was normal to the sample surface. (b) Polar plot of the contrast of visible fluorescence (left) and IR absorption (right) as a function of light-polarization angle for both geometries (see text). Solid lines are fits to a constant and dashed lines are fits to $\sin^2(\theta-\phi_{NV})+c$, where $\theta$ is the polarizer angle, $\phi_{NV}$ is the angle of $\mathbf{z}$ relative to the polarizer axis, and $c$ is an offset which takes into account imperfections in light polarization, beam alignment, and selection rules. For the $\mathbf{z}\perp\mathbf{\hat{k}}$ geometry, the sample and polarizer were positioned such that we expected $\phi_{NV}=90(3)^{\circ}$, and both of the fitted values of $\phi_{NV}$ agree with this value to within the uncertainty.  (c) Light-polarization axes for allowed transitions between the singlet levels \cite{DRE2008}. The selection rules are identical for both absorption and emission.}
\end{figure}

In order to determine the orbital character of the two singlets, we studied the light-polarization selection rules for excitation. A magnetic field was applied which split the resonances belonging to each orientation, allowing us to observe the optical response of each NV orientation (Fig. \ref{fig:pol}(a)). We modified the apparatus illustrated in Fig. \ref{fig:app}(a), inserting separate linear polarizers for both resonant pump (637.7 nm) and probe (1042.4 nm) beams. Figure \ref{fig:pol}(b) shows the light-polarization-dependence of the magnetic-resonance contrast under two geometric configurations; where the NV axis, $\mathbf{z}$, was normal to or lying in the plane of polarization, defined by the normal vector, $\mathbf{\hat{k}}$. For $\mathbf{z}\parallel\mathbf{\hat{k}}$, the contrast of visible fluorescence, $R_{vis}$, is independent of pump polarization angle, and for $\mathbf{z}\perp\mathbf{\hat{k}}$, $R_{vis}$ approaches zero for polarization along $\mathbf{z}$. This behavior is consistent with the expected polarization selection rules for a $E\leftrightarrow A_2$ transition \cite{EPS2005,ALE2007,HOS2008}, where excitation with $\mathbf{x}$ and $\mathbf{y}$ polarized light is equally allowed and $\mathbf{z}$ polarization is forbidden. With pump polarization now held constant at an angle which excited all orientations, we observed the contrast of IR absorption, $R_{ir}$, with varying probe polarization. The results demonstrate that $R_{ir}$ has a similar polarization dependence to $R_{vis}$, indicating that both transitions obey the same polarization selection rules. This observation confirms that the 1042 nm transition involves $E$ and $A_1$ orbitals, as the other possible combinations of singlet levels have different polarization selection rules \cite{DRE2008} (Fig. \ref{fig:pol}(c)).

We have measured optical properties of the 1042 nm transition of the NV center and determined that the transition plays an important role in the optical pumping mechanism. This knowledge can be used to improve NV-based applications; for example, 1042 nm absorption was recently used in a magnetometer which demonstrated about an order of magnitude improvement in sensitivity compared to fluorescence-based techniques \cite{ACO2010APL}. Further work is necessary to determine the relative energies of all four singlet levels and to understand the role of non-radiative decay.

The authors are grateful to A. Gali, N. Manson, L. Rogers, M. Doherty, P. Hemmer, E. Corsini, B. Patton, M. Ledbetter, and L. Zipp for valuable discussions. This work was supported by NSF grant PHY-0855552.

\bibliographystyle{prsty}


\begin{thebibliography}{42}
\bibitem[{IRN()}]{IRNOTE}
\bibinfo{note}{For both of the samples studied in this work, we find the
  transition to occur at 1042.5(3) nm. In Ref. [26], the authors reported the
  ZPL to be at 1046 nm, but, according to private communication with the
  authors, this was a misprint.}
\bibitem[{VIS()}]{VISNOTE2010}
\bibinfo{note}{Because measurements were made in high-defect-density samples,
  where each NV center experiences a different local environment, we neglect
  the difference in lifetimes between $m_s=\pm1$ and $m_s=0$ sublevels.}
\end{thebibliography}

\end{document}